# Cost Efficient Design of Reversible Adder Circuits for Low Power Applications

Neeraj Kumar Misra
Institute of Engineering and Technology, Lucknow, UP

Mukesh Kumar Kushwaha
Institute of Engineering and Technology, Lucknow, UP

Subodh Wairya
Institute of Engineering and Technology, Lucknow, UP

Amit Kumar
Institute of Engineering and Technology, Lucknow, UP

## ABSTRACT
A large amount of research is currently going on in the field of reversible logic, which have low heat dissipation, low power consumption, which is the main factor to apply reversible in digital VLSI circuit design.This paper introduces reversible gate named as 'Inventive0 gate'. The novel gate is synthesis the efficient adder modules with minimum garbage output and gate count. The Inventive0 gate capable of implementing a 4-bit ripple carry adder and carry skip adders. It is presented that Inventive0 gate is much more efficient and optimized approach as compared to their existing design, in terms of gate count, garbage outputs and constant inputs. In addition, some popular available reversible gates are implemented in the MOS transistor design the implementation kept in mind for minimum MOS transistor count and are completely reversible in behaviour more precise forward and backward computation. Lesser architectural complexity show that the novel designs are compact, fast as well as low power.

## 1. INTRODUCTION
Power and speed is an important term in low power VLSI circuit design [4, 6, 7, 10, 12]. There are some conventional methods to optimize the power and speed such as reducing switching activity, lower power supply and reducing technology of the devices. These methods are not fulfilling the criteria to meet present scenario of power optimization. The classical digital approach has been used the digital logic gates, which are irreversible in behaviour. These irreversible digital gates generate energy due to the bit loss during computation. Bit loss occurs because the input and output vector are not equal in number. Thus, classical digital logic dissipates heat for every bit loss during computation (Landauer 1961) more precisely bit loss dissipate $kT \ln 2$ joule of energy where k is Boltzmann's constant and T is the temperature at which computation operation performed [2]. According to Bennett in 1973 proves that in order to nullify the energy loss it is essential that all logic computation operate to be in a reversible logic way [1]. Thus, every latest technology has to use reversible gates in order to power and delay optimization [13, 14, 15, 17].

The paper is categories with the following sections: Section 2 shows the basic reversible gate with its MOS transistor implementation and forward and backward computation logic. The novel reversible gate is introduced which are used in ripple carry adder/subtraction circuits and carry skip adder (In section and subsections 3, 3.1 and 3.2). Section 4 introduces MOS transistor implementation of a proposed Inventive0 gate and simulation result. Section 5 shows the existing methodology and comparison analysis. Finally, concluded with section 6.

## 2. BASIC REVERSIBLE GATE AND ITS MOS TRANSISTOR IMPLEMENTATION
There are various reversible gates existing in the literature, popular gates are FG, FRG and TG [13, 15]. In this section, we design these popular gate in MOS transistor, CAD tools using Microwind DSCH-2.7 is used as circuit design and simulator and transistor channel width and length selected (W=2u,L=0.12u for PMOS and W=1u,L=0.12u for NMOS) for all circuit designs. Because it has a first extraction which Netlist (DSCH-2.7) and Verilog complication (Microwind 2) and timing diagram (DSCH- 2.7)

**(i)** Feynman gate is a 2x2 reversible gate (Fig 1 a). The input vector $I_v$ (A, B) is mapped to output vector $O_V$ (A, A $\oplus$ B) .Design circuit, 8 MOS transistors are required to design FG reversibly (Fig 1 b). The corresponding timing diagram shown in Fig 1 c.

**For forward computation**

P=A; If A=0 then Q= B, else case $Q = \overline{B}$

**For reverse computation**

A=P; If P=0 then B=Q else case $B = \overline{Q}$

**(ii)** Fredkin gate is a 3x3 reversible gate (Fig 2 a).The input vector $I_v$ (A, B, C) is mapped to output vector $O_V$ $(A, \overline{A}B + AC, AB + \overline{A}C)$ .Design circuit, 4 MOS transistors are required to design FRG reversibly (Fig 2 b). The corresponding timing diagram shown in Fig 2 c.

**For forward computation**
P=A; if A=0 then Q= B and R=C else case Q=C and R=B
**For reverse computation**
A=P; if P=0 then B=Q and C=R else case C=Q and B=R.

**(iii)** Toffoli gate is a 3x3 reversible gate (Fig 3 a).The input vector $I_v$ (A, B, C) is mapped to output vector $O_V$ $(A, B, AB \oplus C)$ .Design circuit, 6 MOS transistors are required to design TG reversibly (Fig 3 b). The corresponding timing diagram shown in Fig 3 c.
**For forward computation**

P=A, Q=B, if (A and B) =0 then R=C, else case $R = \overline{C}$
**For reverse computation**
A=P, B=Q, if (P and Q) =0 then C=R else case $C = \overline{R}$

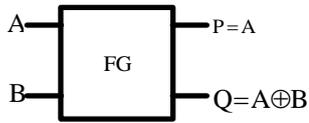

**Fig 1 (a): Block diagram of Feynman Gate.**

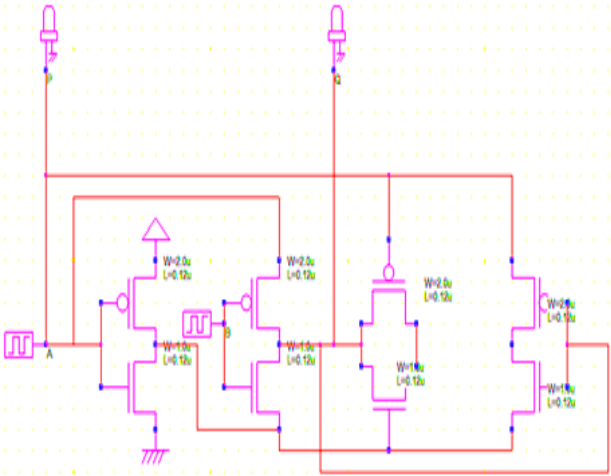

**Fig 1 (b): MOS transistor implementation of FG.**

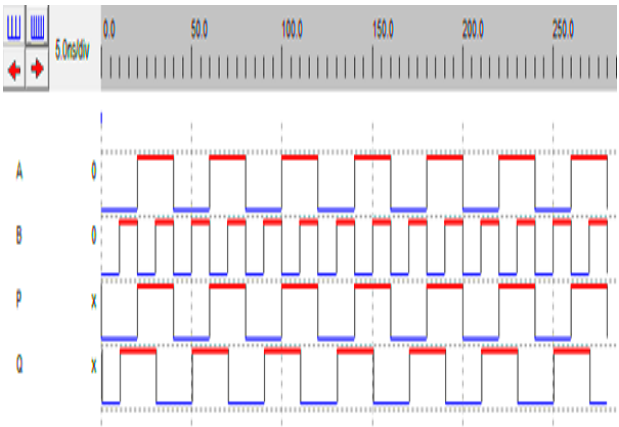

**Fig 1 (c): Simulation result of FG.**

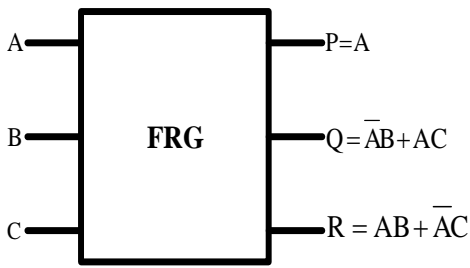

**Fig 2 (a): Fredkin Gate.**

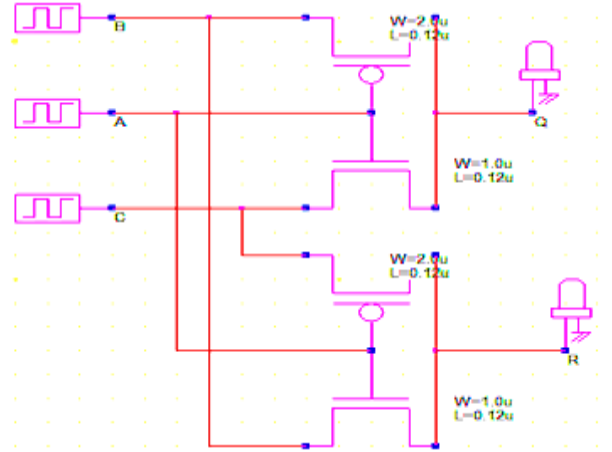

**Fig 2 (b): Fredkin Gate.**

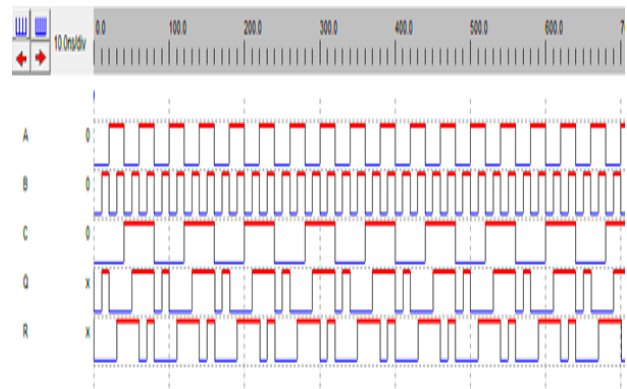

**Fig 2 (c): Simulation result of Fredkin gate.**

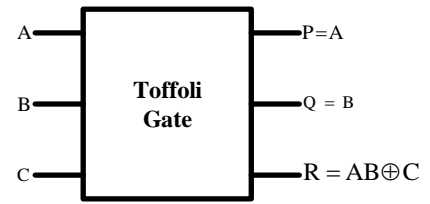

**Fig 3 (a): Toffoli gate**

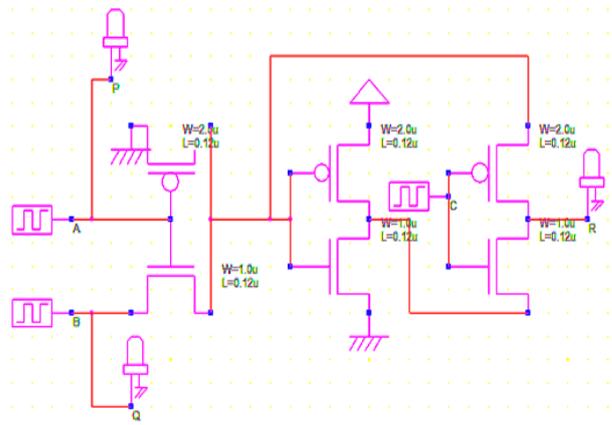

**Fig 3(b): MOS transistor implementation of Toffoli Gate.**

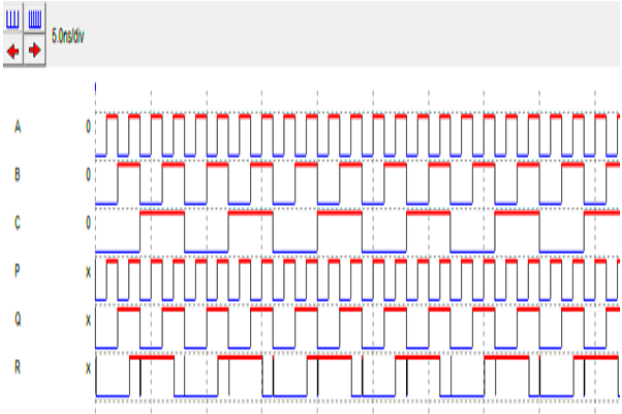

**Fig 3(c): Toffoli gate simulation result.**

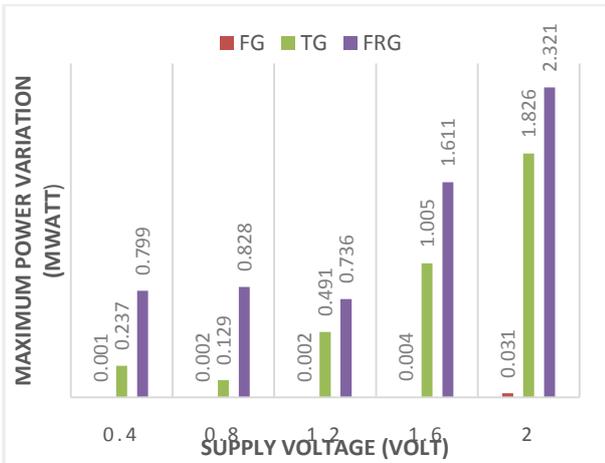

**Fig 4: Power Comparison result at different input voltage.**

## 3. RELATED WORK

In this section, we have designed a new reversible gate named as an Inventive0 gate (Fig 5) that are utilized in ripple-carry adder and carry-skip adder design circuits. Inventive0 gate is a 4x4 gate in order to verify reversibility that the corresponding truth table that the input logic vector and output logic vector are one-to-one correspondence (Table 1). The utility of this gate is that it can operate individually as a reversible full adder as well as reversible full subtraction and it requires lesser hardware complexity also the design cost low. If set input D=0, the corresponding outputs R and S form the full adder. Similarly set input D=1, the corresponding outputs R and S form the full subtraction. Its implementation as a full adder as well as full subtraction are shown in Fig. 6.

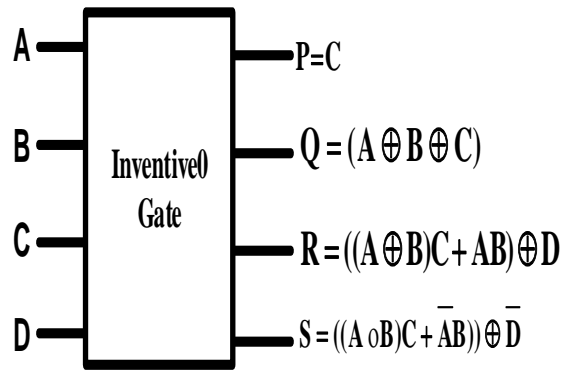

$P = C$

$Q = (A \oplus B \oplus C)$

$R = ((A \oplus B)C + AB) \oplus D$

$S = ((A \circ B)C + \overline{A}B)) \oplus \overline{D}$

**Fig 5. Novel design of 4x4 reversible Inventive0 gate.**

**Table 1. Truth table of the Reversible Inventive0 Gate.**

| Inputs | | | | Outputs | | | |
|---|---|---|---|---|---|---|---|
| **A** | **B** | **C** | **D** | **P** | **Q** | **R** | **S** |
| 0 | 0 | 0 | 0 | 0 | 0 | 0 | 1 |
| 0 | 0 | 0 | 1 | 0 | 0 | 1 | 0 |
| 0 | 0 | 1 | 0 | 1 | 1 | 0 | 0 |
| 0 | 0 | 1 | 1 | 1 | 1 | 1 | 1 |
| 0 | 1 | 0 | 0 | 0 | 1 | 0 | 0 |
| 0 | 1 | 0 | 1 | 0 | 1 | 1 | 1 |
| 0 | 1 | 1 | 0 | 1 | 0 | 1 | 0 |
| 0 | 1 | 1 | 1 | 1 | 0 | 0 | 1 |
| 1 | 0 | 0 | 0 | 0 | 1 | 0 | 1 |
| 1 | 0 | 0 | 1 | 0 | 1 | 1 | 0 |
| 1 | 0 | 1 | 0 | 1 | 0 | 1 | 1 |
| 1 | 0 | 1 | 1 | 1 | 0 | 0 | 0 |
| 1 | 1 | 0 | 0 | 0 | 0 | 1 | 1 |
| 1 | 1 | 0 | 1 | 0 | 0 | 0 | 0 |
| 1 | 1 | 1 | 0 | 1 | 1 | 1 | 0 |
| 1 | 1 | 1 | 1 | 1 | 1 | 0 | 1 |

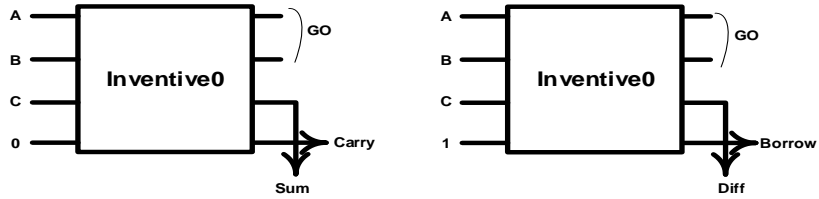

**Fig 6. Inventive0 used as full adder and full subtraction.**

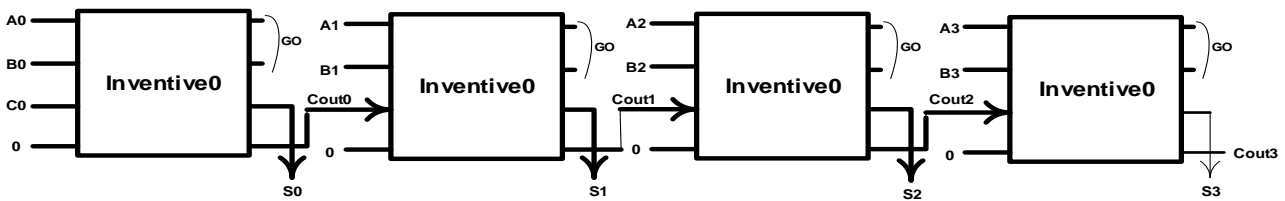

**Fig 7 (a). Inventive0 used as Ripple carry adder.**

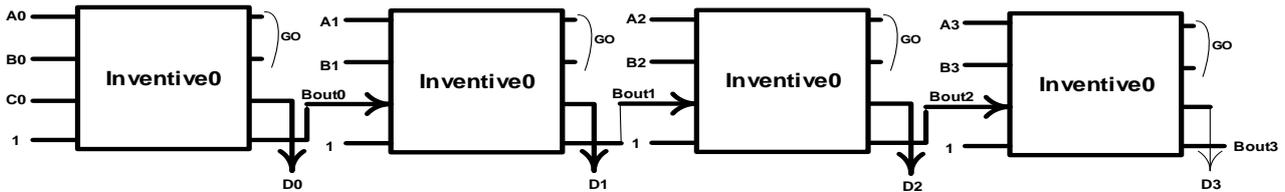

**Fig 7 (b). Inventive0 used as Ripple carry Subtraction.**

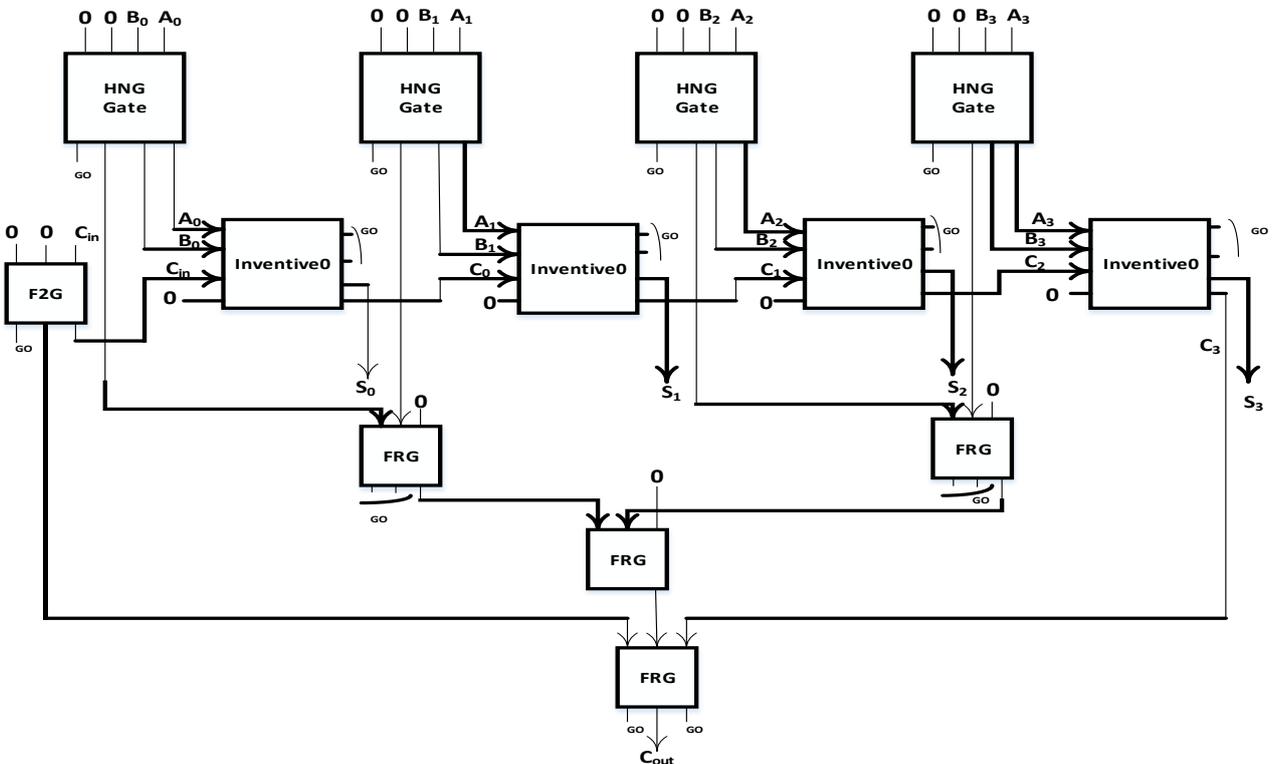

**Fig 8. Novel design of 4-bit Carry skip adder.**

## 3.1 Reversible 4-bit parallel adder/Subtraction

The design utilizes the circuits (Figs 7(a), (b)), as the 4-bit parallel adder/subtraction. In (Figs 7(a), (b)) two 4- bit augends (A=$A_0$ $A_1$ $A_2$ $A_3$ ), addends (B=$B_0$ $B_1$ $B_2$ $B_3$) , $C_{in}$ (Carry in) and $C_{out3}$ (Carry-out).$S_0$-$S_3$ and $D_0$-$D_3$ are the sum or difference outputs depending upon whether the design utilize as ripple carry addition or subtraction.

## 3.2 Novel cost efficient carry-skip adder design methodology

Carry-Skip adder design consist of the following modules: a 4-bit ripple carry adder and carry skip circuit. Carry-Skip circuit basically used to propagate $C_{in}$ (Carry in) to the $C_{out}$ (Carry out) to faster way [5, 17, 18, 19]. In the design circuity $A_i$ and $B_i$ be the four-bit arguments to the $i^{th}$ full-adder. Propagate logic $Pi = Ai \oplus Bi$ are generated, the initial carry-in $C_{in}$ can be immediately associated to the carry-out $C_{out}$. More precisely carry-out $C_{out}$ can be produce without looking to by 4-bit ripple carry adder module. However, in the other fashion, carry-out $C_{out}$ will be produce by 4-bit ripple carry adder module circuit. The carry-out is synthesis by using equation $Cout = \overline{P}C4 \oplus PCin$. In this logic expression if P is high logic, without waiting for the production of final carry ($C_{out}$) other case P is low logic, the carry out $C_{out}$ will be produce by 4-bit ripple carry circuit. The complete block diagram of carry-skip adder as shown in Fig 8.

## 4. MOS REALIZATION OF DIFFERENT ADDER CIRCUIT

The MOS transistor implementation of reversible Invenitve0 gate is proposed (Fig 10). Since in the implementation circuit are based on Gate diffusion input (GDI) for reducing the transistor count.

## 5. PERFORMANCE COMPARSION ANALYSIS

The novel full adder/Subtraction with Inventive0 gate is optimized. In (Fig 9) compares the proposed and existing counterparts. One of the major challenge in reversible technique is to optimize lower bound parameters. We can state that the novel design is optimize parameters.

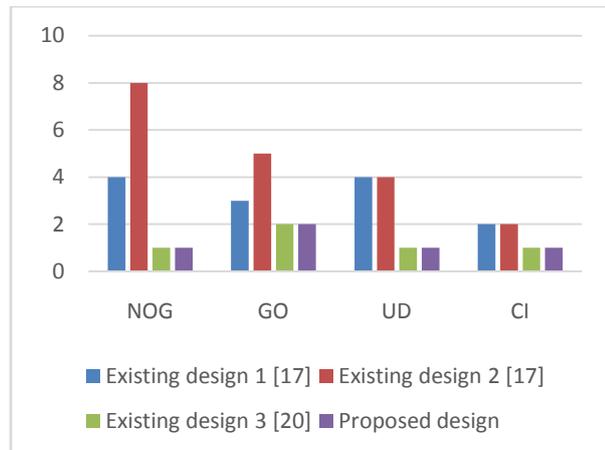

**Fig 9: Parameter analysis of reversible full-adder.**

**Algorithm 1: For construction of carry skip adder circuit.**

**Input:** A= ($A_0$, $A_1$, $A_2$, $A_3$) and B= ($B_0$, $B_1$, $B_2$, $B_3$) and $C_{in}$

**Output:** S= ($S_0$, $S_1$, $S_2$, $S_3$) and $C_{out}$

**Begin**

**Level 1:** For all i in (0, 1, 2, 3)

Compute $Pi = Ai \oplus Bi$ from HNG gate and also copy $A_i$ and $B_i$

**Level 2:** Compute S= ($C_4$, $S_0$, $S_1$, $S_2$, $S_3$) from inventive0 gate. Where $Si = Ai \oplus Bi \oplus Ci$

**Level 3:** Compute carry skip- logic bit $Cout = \overline{P}C4 \oplus PCin$ from FRG gate.

**End;**

**Proposition 1:** A reversible full adder/Subtraction can be realized by at least one reversible gate. It is easy to find that if we set input D=0 and 1 alternatively and then other input (A, B, C) can be added. The output $Q = A \oplus B \oplus C$ and $R = (A \oplus B)C + AB$ for full adder and similarly for $Q = A \oplus B \oplus C$ and $S = (A \, o \, B)C + \overline{A}B$ for full subtraction. Hence, a reversible full adder/Subtraction can be realised by at least one gate.

**Proposition 2:** A 4 reversible gates are needed to implement a 4-bit parallel adder. Since one invetive0 gate can synthesis a reversible full adder. Since 4-bit parallel adder can be design by using cascading 4 full adder (4x Inventive0 gate).Hence at least 4x1 gates should be used to design of 4-bit parallel adder.

**Proposition 3:** A reversible N bit ripple carry adder can be realized by using N number of reversible gates (As Fig 6) .

**Proposition 4:** A reversible N bit ripple carry adder can be realized by using 2N number of garbage output (As Fig 6).

**Proposition 5:** A reversible N bit ripple carry adder can be realized by using N number of constant input (As Fig 6).

**Proposition 6:** A reversible N bit ripple carry adder can be realized by using N number of unit delay (As Fig 6).

**Proposition 7:** A reversible 4-bit ripple carry adder can be realized by following total logical calculation.
$T_{Inventive0}$ = (2α) (for Q-expression) + (2α+ 2β+6δ) (for R-expression) + (1α+2β+2δ) (for S-expression) = 5α+4β+8δ.
Total logical calculation for 4-bit ripple carry adder found by using this expression
TLC (Total logic calculation) = 4 x ($T_{Inventive0}$)
= 4(5α+4β+8δ) = 20α+16β+32δ (As Fig 7(b))
Total logical calculations (circuit cost), One of the primary factors of a circuit is its architectural complexity. It is basically count of the 2 input XOR (denoted by α), 2 input AND (denoted by β), NOT (denoted by δ).

**Proposition 8:** A reversible 4- bit carry adder can be realized by following total logical calculation (T).
In the circuit of 4-bit carry skip adder we can use F2G for fan-out. In the carry skip adder, the TLC (T) for HNG is 5α+2β, T for FRG is (2α+4β+2δ) (As Fig 2a), T for F2G is (2α) .
Therefore, our 4-bit carry skip adder the total logical calculation is:
$T_{carry\ skip}$ = 4x (HNG) +4x (Inventive0 gate) +4x (FRG) + 1x F2G
= 4x (5α+2β) + 4x (5α+4β+8δ) + 4x (2α+4β+2δ) + 2α
= 50α+40β+40δ (As Fig 8).

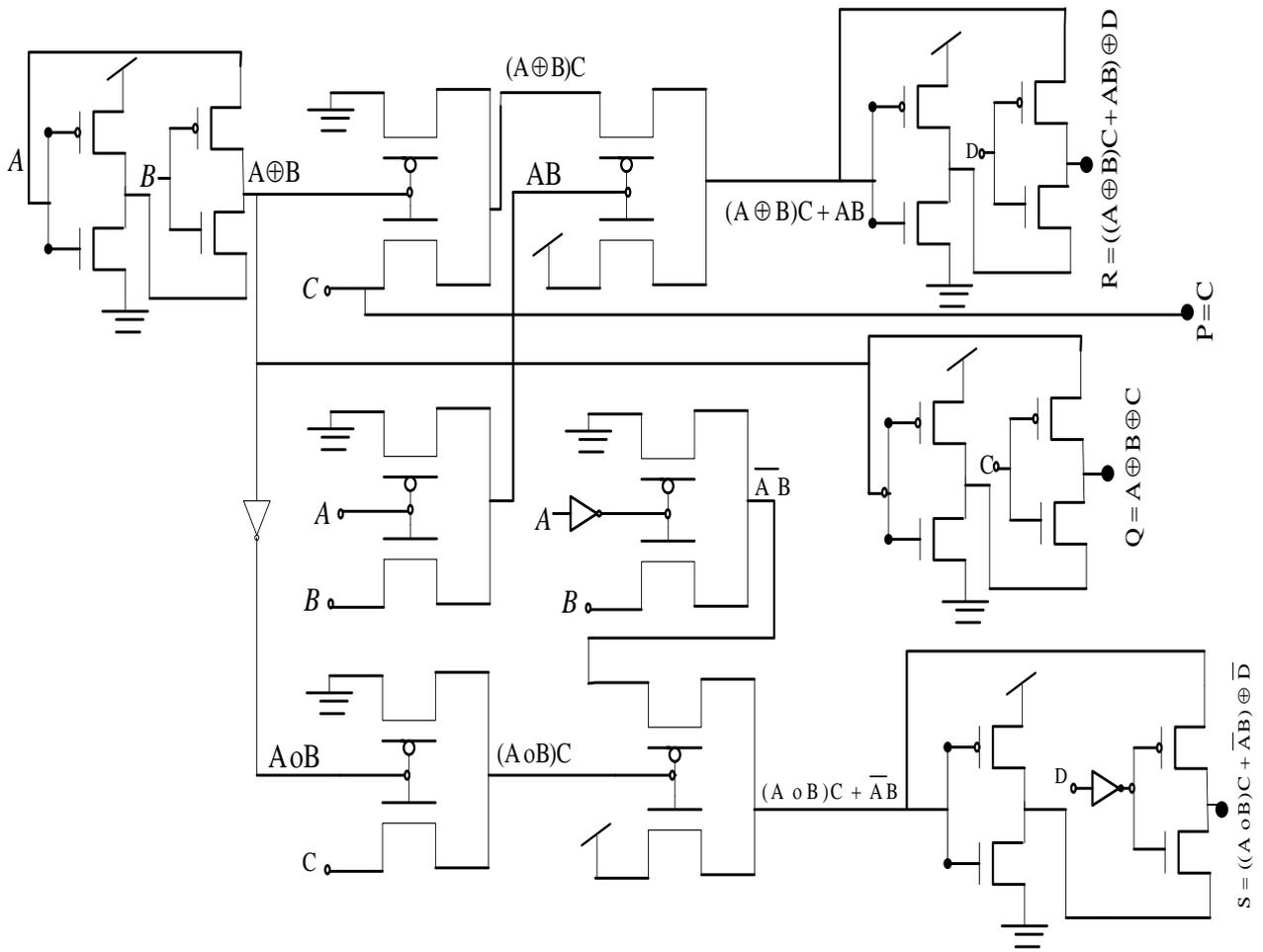

**Fig 10: MOS implementation of Inventive0 gate.**

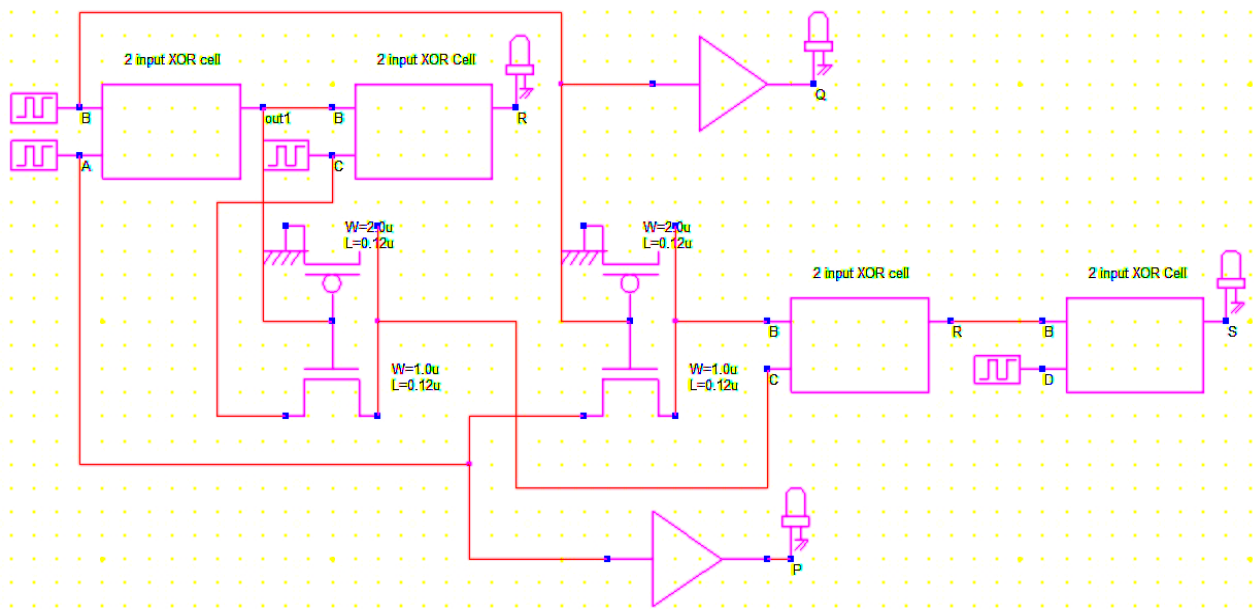

**Fig 11 (a): Circuit implementation of HNG gate.**

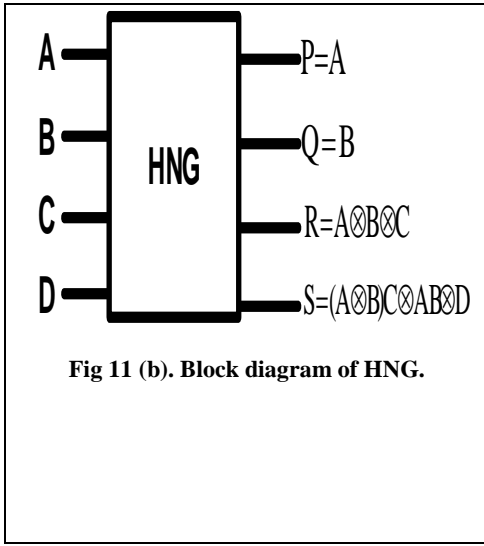

**Fig 11 (b). Block diagram of HNG.**

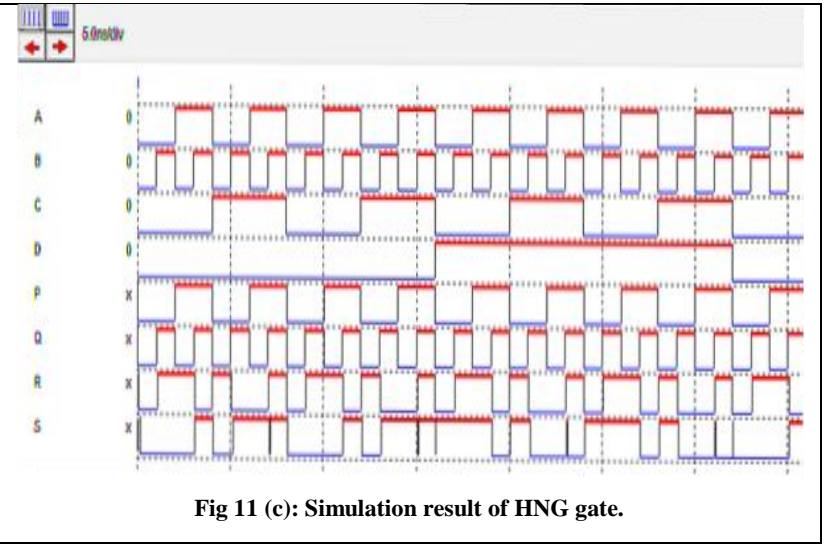

**Fig 11 (c): Simulation result of HNG gate.**

Implementation of Inventive0 gate in Microwind is exhibited in Fig 12(a), which was realized by the minimum MOS transistor count.

If Inventive0 gate form a Full adder setting D input ON and OFF timing equal to Zero. A, B and C inputs are in bit form, then it works as a full adder shown in Fig 12 (b).The corresponding truth table of Inventive0 as full adder in Table 2.

If Inventive0 gate form a Full subtraction setting D input ON and OFF timing equal to one. A, B and C inputs are in bit form, then it works as a full subtraction shown in Fig 12 (c).The corresponding truth table of Inventive0 as full subtraction in Table 3.

**Table 2. Truth table of Inventive0 as full adder.**

| A | B | C | Sum $= A \oplus B \oplus C$ | Carry $= (A \oplus B)C + AB$ |
|---|---|---|---|---|
| 0 | 0 | 0 | 0 | 0 |
| 0 | 0 | 1 | 1 | 0 |
| 0 | 1 | 0 | 1 | 0 |
| 0 | 1 | 1 | 0 | 1 |
| 1 | 0 | 0 | 1 | 0 |
| 1 | 0 | 1 | 0 | 1 |
| 1 | 1 | 0 | 0 | 1 |
| 1 | 1 | 1 | 1 | 1 |

**Table 3. Truth table of Inventive0 as full subtraction.**

| A | B | C | Difference $= A \oplus B \oplus C$ | Borrow $= (A \circ B)C + \overline{A}B$ |
|---|---|---|---|---|
| 0 | 0 | 0 | 0 | 0 |
| 0 | 0 | 1 | 1 | 1 |
| 0 | 1 | 0 | 1 | 1 |
| 0 | 1 | 1 | 0 | 1 |
| 1 | 0 | 0 | 1 | 0 |
| 1 | 0 | 1 | 0 | 0 |
| **1** | 1 | 0 | 0 | 0 |
| **1** | 1 | 1 | 1 | 1 |

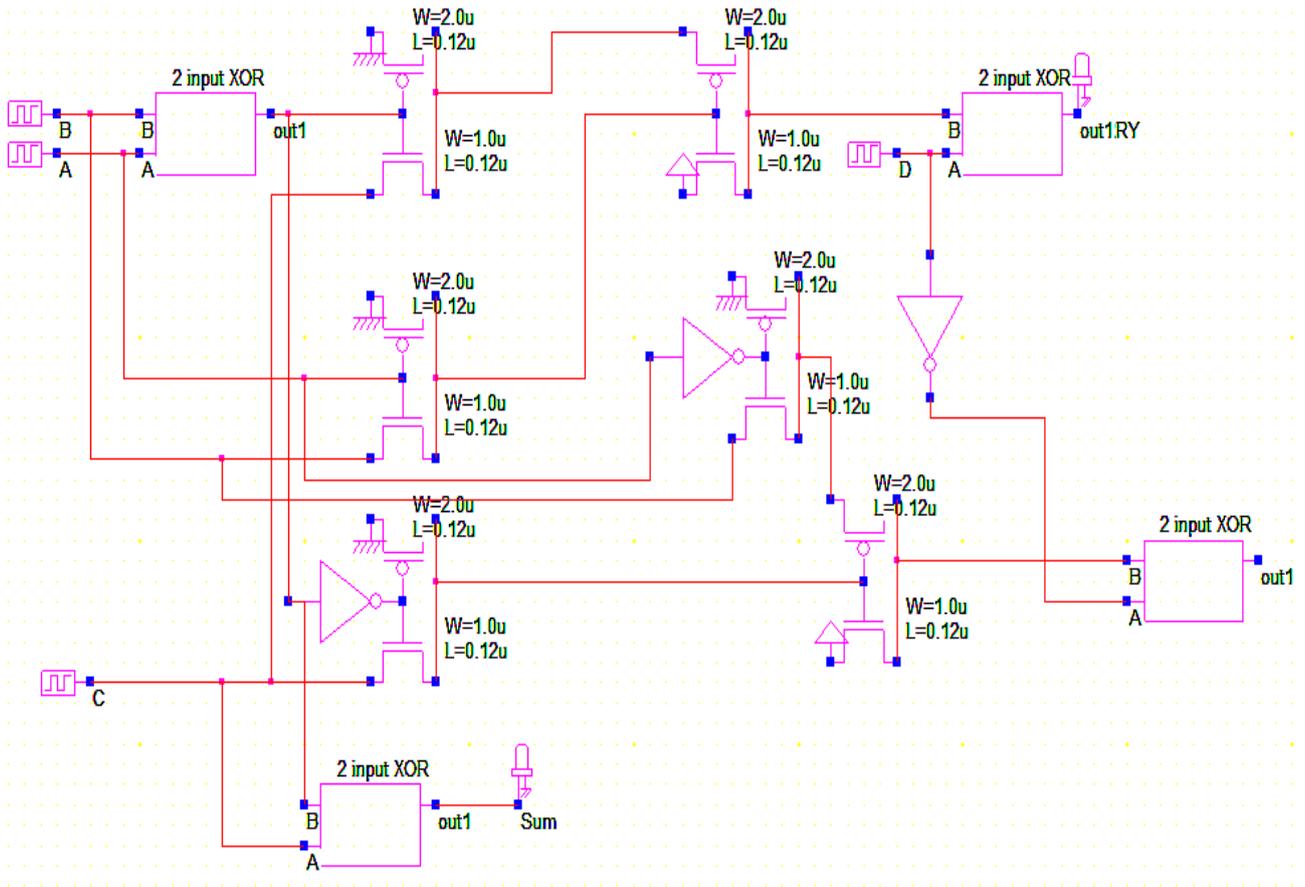

**Fig 12 (a): Circuit implementation of Inventive0 gate.**

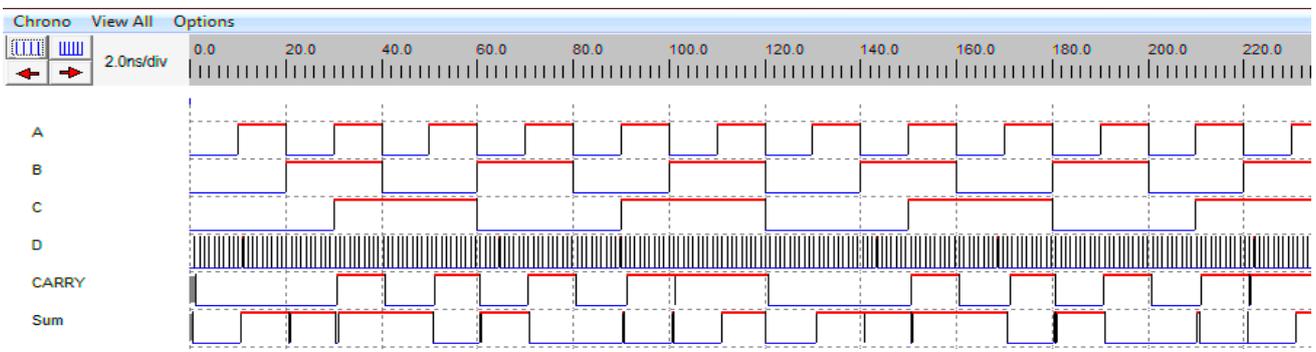

**Fig 12 (b): Simulation result of Invenitve0 gate as Full adder.**

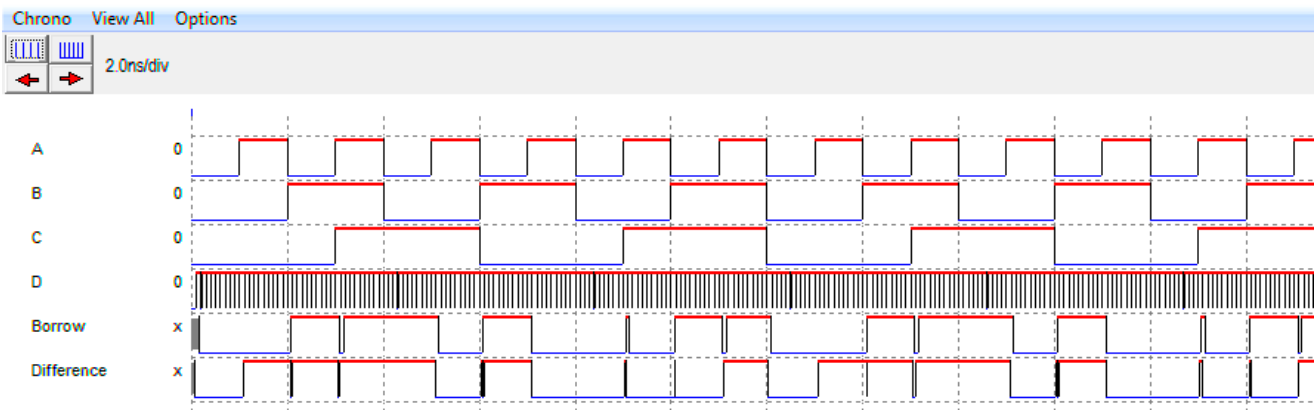

**Fig 12 (c): Simulation result of Invenitve0 gate as Full subtraction.**

## 6. CONCLUSION
The centers of this paper are the set of operations, including full adder/Subtraction, ripple carry adder/Subtraction and carry skip adder using a novel 4x4 reversible Inventive0 gate. It has also been shown by the comparative behavior that the novel designs has utilize with the optimize parameters such as garbage output, constant input and total logical calculation. In addition the novel Inventive0 gate implemented in MOS transistor using the gate diffusion input technique with the minimum transistor count. Since addition and subtraction are useful in many operations such as ALU, Digital signal processing, communication systems and low power circuits etc.

## 7. REFERENCES

[1] C.H. Bennett. 1973. Logical reversibility of Computation, IBM Journal Research and Development. 17, 525-532.

[2] R. Landauer. 1961. Irreversibility and Heat Generation in the Computing Process. IBM Journal of Research and Development, 5(3), 183-191.

[3] Diganta Sengupta, Mahamuda Sultana, Atal Chaudhuri. 2011. Relization of a Novel Reversible SCG Gate and its Application for designing Parallel adder/subtractor and match logic. International Journal of Computer Applications, v 31(9), 30-34.

[4] Neeraj Kumar Misra, Subodh Wairya and Vinod Kumar Singh. 2014. An Inventive Design of 4*4 Bit Reversible NS Gate. IEEE International Conference on Recent Advances and Innovation in Engineering (ICRAIE-2014), 1-6.

[5] Himanshu Thapliyal and Nagarajan Ranganathan. 2011. A New reversible design of BCD adder. In proceeding of the design Automation and test in Europe (DATE). Grenoble. France. 1180-1183.

[6] Neeraj Kumar Misra, Subodh Wairya and Vinod Kumar Singh. 2014. Preternatural Low-Power Reversible Decoder Design in 90 nm Technology Node. International Journal of Scientific & Engineering Research. 5(6). 969-978.

[7] Subodh Wairya, Rajendra Kumar Nagaria and Sudarshan Tiwari. 2012. Performance Analysis of High Speed Hybrid CMOS Full Adder Circuits for Low Voltage VLSI Design, Hindawi Publishing Corporation VLSI Design.

[8] Michael P. Frank. 2005. Introduction to Reversible Computing: Motivation, Progress, and Challenges, Proceedings of the 2nd conference on computing frontier Ischia, Italy, 385-390.

[9] Neeraj Kumar Misra, Subodh Wairya and Vinod Kumar Singh, 2014. Evolution of structure of some binary group-based n-bit comparator, n-to-$2^n$ decoder by reversible technique, International Journal of VLSI design & Communication Systems (VLSICS), v 5(5), 9-22

[10] Neeraj Kumar Misra, Subodh Wairya and Vinod Kumar Singh. 2015. Approaches to Design Feasible Error Control Scheme Based on Reversible Series Gates. European Journal of Scientific Research. vol. 129, no. 3, 224-240.

[11] Md. M. H Azad Khan, 2002. Design of Full-adder With Reversible Gates. International Conference on Computer and Information Technology, Dhaka, Bangladesh, 515-519.

[12] Neeraj Kumar Mishra et al., 2013. An advancement in the NxN Multiplier Architecture Realization via the Ancient indian Vedic Mathematics, International Journal of Electronics Communication and Computer Engineering, vol (4), Issue(2), 2278-4209.

[13] Bibhash Sen, et al,. 2013. Reversible Logic-Based Fault-Tolerant Nanocircuits in QCA, ISRN Electronics Hindawi Publishing Corporation.

[14] Hafiz Md. Hasan babu, Nazir Saleheen, Lafifa Jamal, Sheikh Muhammad Sarwar, 2013, Approach to design a compact reversible low power binary comparator, IET Computers and Digital Technique, vol (8), Issue. 3, 129-139.

[15] H. Thapliyal and M. Srinivas. 2005. Novel Reversible 'TSG' Gate and Its Application for Designing Components of Primitive Reversible/Quantum ALU, Fifth International Conference on Information, Communications and Signal Processing, Bangkok, 1425-1429

[16] Rangaraju H G, Venugopal U, Muralidhara K N, Raja K B. 2010. Low Power Reversible Parallel Binary Adder / Subtractor, International journal of VLSI design & Communication Systems (VLSICS). vol.1, no.3, 1-12.

[17] Biswas, A. K., Hasan, M. M., Hasan, M., Chowdhury, A. R., & Hasan Babu. 2008. A novel approach to design BCD adder and Carry Skip BCD adder, 21st International Conference on VLSI Design, 566-571.

[18] Zhou, Rigui, Manqun Zhang, Qian Wu, and Yang Shi. 2012. Designing novel reversible BCD adder and parallel adder/subtraction using new reversible logic gates, International Journal of Electronics. v (99), no. 10, 1395-1414.

[19] Krishnaveni, D., and Geetha Priya, M. 2010, A Novel Design of Reversible Serial and Parallel Adder/Subtractor, International Journal of Engineering Science and Technology, vol 3, 2280–2288.